# Optically detected nuclear quadrupolar interaction of [14]N in nitrogen-vacancy centers in diamond

Chang S. Shin[1,2,*], Mark C. Butler[1,2,†], Hai-Jing Wang[1,2], Claudia E. Avalos[1,2], Scott J. Seltzer[1,2], Ren-Bao Liu[3], Alexander Pines[1,2], and Vikram S. Bajaj[1,2,‡]

[1]*Materials Sciences Division, Lawrence Berkeley National Laboratory, Berkeley, California 94720, USA*

[2]*Department of Chemistry and California Institute for Quantitative Biosciences, University of California, Berkeley, California 94720, USA*

[3]*Department of Physics and Centre for Quantum Coherence, The Chinese University of Hong Kong, Shatin, New Territories, Hong Kong, China*

## ABSTRACT

We report sensitive detection of the nuclear quadrupolar interaction of the [14]N nuclear spin of the nitrogen-vacancy (NV) center using the electron spin echo envelope modulation technique. We applied a weak transverse magnetic field to the spin system so that certain forbidden transitions became weakly allowed due to second-order effects involving the nonsecular terms of the hyperfine interaction. The weak transitions cause modulation of the electron spin-echo signal, and a theoretical analysis suggests that the modulation frequency is primarily determined by the nuclear quadrupolar frequency; numerical simulations confirm the analytical results and show excellent quantitative agreement with experiments. This is an experimentally simple method of detecting quadrupolar interactions, and it can be used to study spin systems with an energy structure similar to that of the nitrogen vacancy center.

# I.   INTRODUCTION

Recently, NV centers in diamond have drawn much attention for applications ranging from physics to biology, owing to their favorable optical and spin properties. High spin polarization to the $m_s = 0$ state at room temperature via optical pumping,[1] convenient optical read-out of the spin states via spin-state dependent fluorescence detection, and long electron-spin coherence time of milliseconds[2] at room temperature offer opportunities for using NV centers as a sensitive detector or in spin-based quantum information technologies.[3–10] Here we report the detection of the nuclear quadrupolar interaction of the [14]N nuclear spin associated with NV centers.

The interaction of nuclei with the local electric field gradient can provide information about orbital electron states, and so the nuclear quadrupole coupling constant is often used to study bond hybridization, degree of covalency, and orbital population in molecules that have a nucleus of angular momentum $I \geq 1$.[11,12] It can also be used as a fingerprint of target molecules in narcotics and explosives,[13,14] due to the strong dependence of the quadrupole coupling constant on the electronic environment.[11,15–17] Several techniques are used to measure nuclear quadrupolar interactions. For instance, nuclear magnetic resonance (NMR) spectroscopy is used to measure the quadrupolar coupling   in cases where it is a small perturbation to the much larger Zeeman interaction.[12] For atoms with a large atomic number, however, the quadrupolar coupling can be comparable to or larger than the nuclear Zeeman interaction inside a high-field NMR spectrometer. In this case, nuclear quadrupole resonance (NQR) in low magnetic field or near-zero magnetic field[18–20] can be used to measure "pure quadrupole resonance."   In NQR, a RF excitation pulse is applied at a frequency resonant with a transition of the quadrupolar Hamiltonian, resulting in a linearly oscillating FID-like signal. However, at near-zero magnetic fields, the thermal polarization at room temperature is very low because $H_Q/kT$ is small. NQR measurements therefore often require ultrasensitive detectors, such as superconducting quantum interference devices[21] or vapor cell magnetometers.[22]

Alternatively, electron spin echo envelope modulation (ESEEM)[23] has been used on photo-excited triplet states of certain molecular crystals to measure nuclear quadrupole coupling constants and/or hyperfine coupling constants.[19,24,25] In a magnetic field applied parallel to the electron-spin quantization axis, the anisotropic terms of the hyperfine interaction, such as $S_z A_{zx} I_x$, give different nuclear-spin eigenstates for different states of the electron spin. A transition of the electron spin projects the nuclear spin onto a different set of eigenstates, and as a result, the envelope of the electron spin echo is modulated at frequencies determined by the hyperfine interaction, the nuclear quadrupolar interaction, and the nuclear Zeeman interaction.[26]

The $^{14}$N nuclear spin associated with NV center, however, does not induce any modulation in the electron spin-echo envelope when a magnetic field is applied parallel to the NV quantization axis. This is because the quantization axis of the $^{14}$N nuclear spin is parallel to the NV quantization axis, and the nuclear spin eigenstates are independent of the electron spin state, so any effect of the frequency shifts due to interactions involving the nuclear spin is completely removed by the spin echo. Other more elaborate techniques, such as optically detected Raman heterodyne NMR or optically detected Raman heterodyne electron nuclear double resonance (ENDOR) have therefore been employed to extract the nuclear quadrupole coupling constant from the dependence of the spectral peaks on magnetic field strength (~1000 G) and orientation.[27,28]

For the experiments reported here, we applied a small transverse magnetic field perpendicular to the NV quantization axis and observed modulations in the echo envelope of the NV electron spin, as shown in Fig. 1. The theoretical model described below shows that the transverse magnetic field mixes electron spin states, which in turn leads to mixing between product states of the electron and nuclear spin due to the hyperfine interaction terms $S_+I_-$ and $S_-I_+$. This mixing of product states, which is a second-order effect involving two nonsecular terms in the Hamiltonian, causes certain forbidden transitions to become weakly allowed. In general, forbidden transitions lead to modulation of the electron spin-echo envelope,[23] and the simulations described below confirm that under our experimental conditions, the nuclear quadrupolar Hamiltonian determines the modulation frequencies.

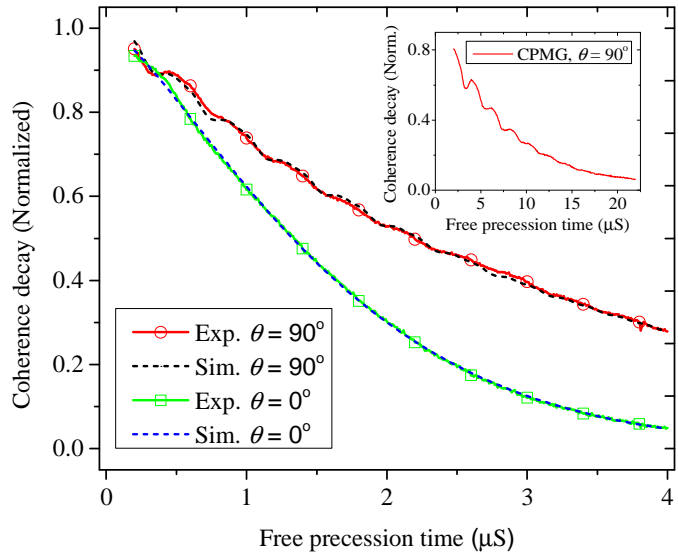

**FIG. 1**. (Color online) Optically detected ESEEM of the NV center in diamond. Solid lines represent experimental results (circle or square markers are used only for the purpose of figure legends), and dashed lines represent simulation results. An external magnetic field of 75 G was applied along the NV axis ($\theta = 0^o$) or perpendicular to the NV axis ($\theta = 90^o$). Modulation of the exponential decay for $\theta = 90^o$ is due to the $^{14}$N nuclear quadrupolar coupling of the NV center. The inset shows the experimental result for the CPMG pulse sequence with ten $\pi$ pulses when the magnetic field is orientated at $\theta = 90^\circ$.

## II.   ODESR EXPERIMENTS WITH THE NV CENTER

Optically detected electron spin resonance (ODESR) was employed to measure the coherence decay of the NV center using the Hahn echo pulse sequence[23,29] or the Carr-Purcell-Meiboom-Gill (CPMG) pulse sequence[30,31] in the presence of a static magnetic field of ~75 G, applied perpendicular to the NV axis. A diamond sample with an estimated NV concentration of ~ 5 ppm and a nitrogen concentration of < ~100 ppm was mounted on a printed circuit board, and a MW field was applied using a small loop (~1.5 mm in diameter) fabricated on a printed circuit board. The MW frequency was matched to the transition frequency between two electron states (denoted by $|\psi_z\rangle$ and $|\psi_y\rangle$ in Sec. III). The NV center was optically excited by a 532-nm laser, and the fluorescence signal was detected by an avalanche photo diode. The detailed experimental procedure for ODESR is described elsewhere.[5,32]

The NV center in diamond consists of a substitutional nitrogen atom at a carbon lattice site and a vacancy adjacent to the nitrogen atom, as shown in Fig.2. The negatively charged NV center has electron spin $S = 1$ in the ground state with $m_s = 0$ and $m_s = \pm 1$ sub-levels that are separated by a zero-field splitting of 2.87 GHz that characterizes the spin-spin interactions.[33–35]

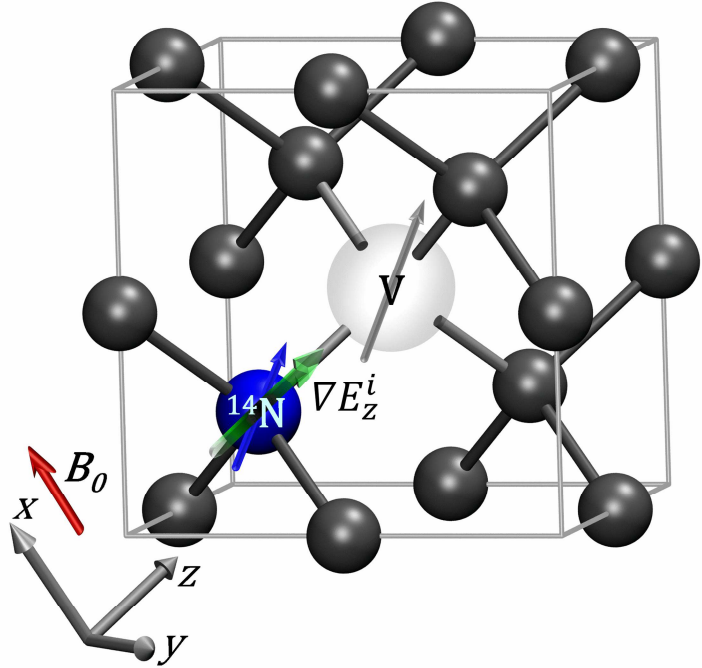

**FIG. 2.** (Color online) Illustration of the negatively charged NV center in diamond. The blue sphere with a blue arrow represents the $^{14}$N nuclear spin, and the light gray sphere with a gray arrow at the vacancy represents the NV electron spin. The green arrow represents the $z$ component of the electric field gradient at the $^{14}$N nuclear site. Dark gray spheres represent carbon atoms in the diamond lattice. The red arrow represents the external magnetic field along the $x$ axis.

ESEEM has previously been studied for NV centers in diamond with natural-abundance (~1.1%) $^{13}$C nuclei and low (of the order of part per billion) $^{14}$N impurities. Under these conditions, the envelope modulation of electron spin-echo signals is frequently dominated by the strongly coupled $^{13}$C nuclear spins that are randomly distributed in the diamond lattice. [36,37] Due to the anisotropic hyperfine interaction between NV centers and neighboring $^{13}$C nuclear spins, the electron spin-echo signal shows modulation even in a magnetic field applied along the NV quantization axis, and the modulation becomes complicated owing to the position-dependent modification of the g-factor for the neighboring $^{13}$C nuclei when misalignment of the magnetic field from the NV axis enhances mixing of electron and nuclear spin states.[37–39]

In contrast to the hyperfine interaction with $^{13}$C, the hyperfine interaction with the $^{14}$N nuclear spin of the NV center is almost isotropic and has principal axes parallel to the NV axes.[40] Nuclear-spin mixing within manifolds defined by a given eigenstate of the electron spin is generally suppressed due to the large quadrupolar interaction (~5 MHz) that is also parallel to the NV quantization axis.[41] Therefore, modulation associated with the $^{14}$N nuclear spin of the NV center has not been observed in the spin-echo envelope.

However, a weak transverse magnetic field (~75 G) can induce mixing in the electron spin states, and we show below that as a result, non-secular terms of the hyperfine interaction enable forbidden transitions that cause modulation of the spin-echo envelope at the $^{14}$N quadrupolar frequency. Although the Zeeman interaction with a weak transverse field is still a small perturbation to the large zero-field-splitting Hamiltonian (~1000 G) of the NV center, the mixing of electron spin states induced by the Zeeman interaction is accompanied by mixing of product states induced by the hyperfine interaction. The forbidden transitions that prevent complete refocusing by a MW $\pi$ pulse in a spin echo are due to this mixing of product states, a second-order effect involving the transverse field as well as the nonsecular terms of the hyperfine interaction.

### III. THEORY

The spin Hamiltonian of the NV center is[33,40]

$$H = H_{\text{ZFS}} + H_{\text{B}} + H_{\text{Q}} + H_{\text{HF}}, \qquad (1)$$

where

$$H_{ZFS} = D\left[S_z^2 - S(S+1)/3\right] + E\left(S_x^2 - S_y^2\right),$$ (1a)

$$H_B = \gamma_e B_0 \cdot S - \gamma_N B_0 \cdot I,$$ (1b)

$$H_Q = P\left[I_z^2 - I(I+1)/3\right],$$ (1c)

$$H_{HF} = S \cdot A \cdot I = H_{HF\parallel} + H_{HF\perp} = A_\parallel S_z I_z + A_\perp\left(S_+ I_- + S_- I_+\right)/2.$$ (1d)

In Eqns. (1a) through (1d), $D/2\pi$ = 2870 MHz and $E/2\pi$ = ~2.75 MHz (sample dependent) are zero-field splitting parameters of the NV center, $P/2\pi$ = -5.04 MHz is the nuclear quadrupole coupling,[40] $\gamma_e$ is gyromagnetic ratio of the NV electron spin, $\gamma_N$ is the gyromagnetic ratio of the $^{14}$N nuclear spin, and $A$ is the hyperfine tensor with $A_\parallel/2\pi$ = 2.3 MHz, $A_\perp/2\pi$ = 2.1 MHz. The energy level diagram for the NV center in the presence of a transverse magnetic field of 75 G along the $x$ axis is shown in Fig. 3.

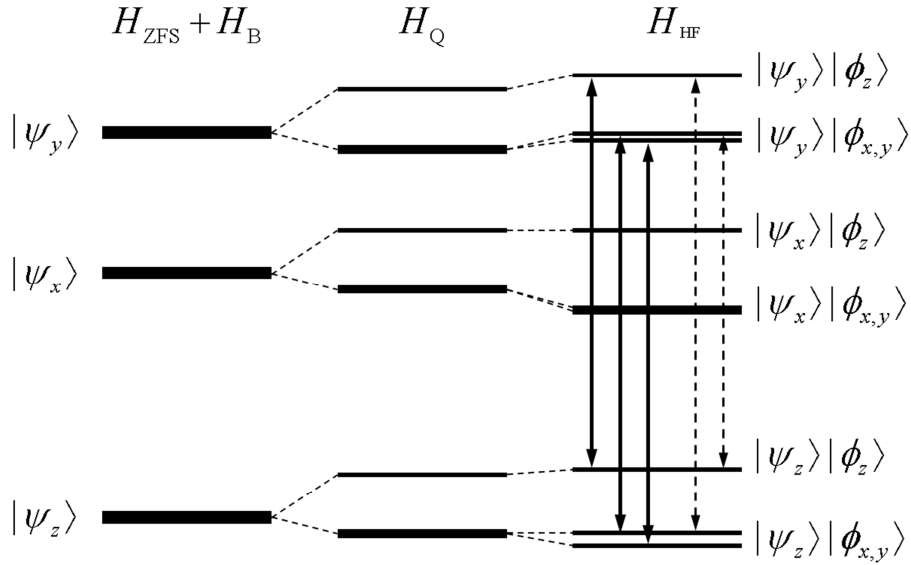

**FIG. 3.** Energy level diagram for the NV center in diamond in the presence of a transverse magnetic field of 75 G along the $x$-axis. The Hamiltonian $H_{ZFS}$ is responsible the zero-field splitting of the NV center, $H_B$ governs the Zeeman interaction, $H_Q$ governs the quadrupolar interaction, and $H_{HF}$ governs the hyperfine interaction. Approximate energy eigenstates are indicated next to the corresponding energy levels, where $|\psi_x\rangle$, $|\psi_y\rangle$ and $|\psi_z\rangle$ are electron spin states, while $|\phi_x\rangle$, $|\phi_y\rangle$ and $|\phi_z\rangle$ are nuclear spin states. On the right side of the figure, the product state making the dominant contribution to each energy eigenstate is shown. Allowed (solid arrows) and forbidden (dashed arrows) spin transitions are also shown.

For qualitative analysis, let us first consider a spin Hamiltonian where the hyperfine interaction $H_{HF}$ has been turned off in Eq. (1d). In the presence of a weak transverse external magnetic field $B_0$ along the $x$-axis, the energy eigenstates of the electron spin are approximately

$$|\psi_z^{(0)}\rangle = |m_s = 0\rangle , \tag{2a}$$

$$|\psi_x^{(0)}\rangle = \tfrac{1}{\sqrt{2}}\left(-|m_s = +1\rangle + |m_s = -1\rangle\right) , \tag{2b}$$

$$|\psi_y^{(0)}\rangle = \tfrac{i}{\sqrt{2}}\left(|m_s = +1\rangle + |m_s = -1\rangle\right) , \tag{2c}$$

and the nuclear eigenstates are approximately

$$|\phi_z^{(0)}\rangle = |m_n = 0\rangle , \tag{2d}$$

$$|\phi_x^{(0)}\rangle = \tfrac{1}{\sqrt{2}}\left(-|m_n = +1\rangle + |m_n = -1\rangle\right) , \tag{2e}$$

$$|\phi_y^{(0)}\rangle = \tfrac{i}{\sqrt{2}}\left(|m_n = +1\rangle + |m_n = -1\rangle\right) . \tag{2f}$$

(Note that states with subscript $x$, $y$, or $z$ are eigenstates of the corresponding Cartesian spin operators.) In Eqns. (2a) through (2f), the superscript indicates that these states are zero-order approximations that do not include mixing of electron spin states separated by the zero-field splitting $D$ or nuclear spin states separated by the quadrupole coupling $P$. The mixing due to the Zeeman interaction, neglected in these zero-order approximations, is represented by coefficients $\sim \gamma_e B_0 / D$ for electron spins and $\sim \gamma_N B_0 / P$ for the nuclear spin.

A transverse magnetic field of ~75 G together with the $E$ splitting separates the energies of the electron spin eigenstates $|\psi_y^{(0)}\rangle$ and $|\psi_x^{(0)}\rangle$ by ~20 MHz. We studied only the transitions associated with states $|\psi_z^{(0)}\rangle$ and $|\psi_y^{(0)}\rangle$ due to a resonant MW $\pi/2$ pulse of 50 ns; for simplicity, our analysis neglects possible off-resonant effects involving the state $|\psi_x^{(0)}\rangle$. In addition to causing shifts in the energy, the transverse field induces small but non-negligible mixing of states $|\psi_z^{(0)}\rangle$ and $|\psi_y^{(0)}\rangle$, such that the electron spin eigenstates can be approximated as $|\psi_z\rangle \approx |\psi_z^{(0)}\rangle + \delta|\psi_y^{(0)}\rangle$ and $|\psi_y\rangle \approx \delta^{'}|\psi_z^{(0)}\rangle + |\psi_y^{(0)}\rangle$ with mixing coefficients $|\delta| \approx |\delta^{'}| \approx \gamma_e B_0/D \sim 0.07$. At the same level of approximation, the nuclear spin eigenstates are $|\phi_z\rangle \approx |\phi_z^{(0)}\rangle$ and $|\phi_{x,y}\rangle \approx |\phi_{x,y}^{(0)}\rangle$, since $\gamma_N B_0/P$ is an order of magnitude smaller than $\gamma_e B_0/D$. The electron spin state $|\psi_x\rangle$ can be identified with $|\psi_x^{(0)}\rangle$, which is an eigenstate of the Hamiltonian representing the Zeeman interaction with the transverse field. Note that in the notation for states $|\psi_y\rangle$ and $|\psi_z\rangle$, we retain the subscripts used for the corresponding zero-order states, in spite of the fact that these mixed electron states are not eigenstates of Cartesian spin operators.

Turning on the term proportional to $S_z I_z$ in the hyperfine interaction causes mixing between product states $|\psi_y\rangle|\phi_x\rangle$ and $|\psi_x\rangle|\phi_y\rangle$, as well as between $|\psi_y\rangle|\phi_y\rangle$ and $|\psi_x\rangle|\phi_x\rangle$. In particular, the states $|\psi_y\rangle|\phi_x\rangle$ and $|\psi_y\rangle|\phi_y\rangle$ that participate in resonant transitions are replaced by $|\psi_y\rangle|\phi_x\rangle + \varepsilon|\psi_x\rangle|\phi_y\rangle$ and $|\psi_y\rangle|\phi_y\rangle + \varepsilon^{'}|\psi_x\rangle|\phi_x\rangle$, respectively, where the coefficients $\varepsilon$ and $\varepsilon^{'}$ represent mixing due to the hyperfine interaction. The forbidden transitions enabled by this mixing cause modulation of the spin-echo envelope. For example, the transition $|\psi_z\rangle|\phi_y\rangle \leftrightarrow |\psi_y\rangle|\phi_x\rangle$ is forbidden by the selection rules for the MW field

because of the change in the nuclear spin state, but the transition $|\psi_z\rangle|\phi_y\rangle \leftrightarrow \left(|\psi_y\rangle|\phi_x\rangle + \varepsilon|\psi_x\rangle|\phi_y\rangle\right)$ is weakly allowed. The transition $|\psi_z\rangle|\phi_y\rangle \leftrightarrow \left(|\psi_y\rangle|\phi_y\rangle + \varepsilon'|\psi_x\rangle|\phi_x\rangle\right)$ is also allowed, and so a MW $\pi/2$ pulse creates a coherence between $|\psi_z\rangle|\phi_y\rangle$ and a linear combination of the states $|\psi_y\rangle|\phi_x\rangle + \varepsilon|\psi_x\rangle|\phi_y\rangle$ and $|\psi_y\rangle|\phi_y\rangle + \varepsilon'|\psi_x\rangle|\phi_x\rangle$. The evolution of this linear combination causes modulation in the spin-echo envelope. Since the nuclear spin states states $|\phi_x\rangle$ and $|\phi_y\rangle$ have the same energy under the quadrupolar Hamiltonian, however, the energy difference between states $|\psi_y\rangle|\phi_x\rangle + \varepsilon|\psi_x\rangle|\phi_y\rangle$ and $|\psi_y\rangle|\phi_y\rangle + \varepsilon'|\psi_x\rangle|\phi_x\rangle$ is small, and the envelope modulation occurs at very low frequencies.

Now let us consider how the eigenstates change when $H_{\text{HF}\perp} = A_\perp\left(S_+I_- + S_-I_+\right)$ is also included in the spin Hamiltonian. In the basis of product states $|\psi_j\rangle|\phi_k\rangle$, the largest matrix elements of $H_{\text{HF}\perp}$ have magnitude $\sim A_\perp$, and these cause mixing between pairs of product states for which the electron is in distinct spin states, e.g. $|\psi_y\rangle|\phi_z\rangle$ and $|\psi_z\rangle|\phi_y\rangle$. The operator $H_{\text{HF}\perp}$ also has smaller matrix elements of magnitude $\sim \gamma_e B_0 \cdot A_\perp/D$ that mix $|\psi_y\rangle|\phi_y\rangle$ and $|\psi_y\rangle|\phi_z\rangle$ as well as $|\psi_z\rangle|\phi_y\rangle$ and $|\psi_z\rangle|\phi_z\rangle$, due to the fact that $|\psi_z\rangle$ and $|\psi_y\rangle$ are each a mixture of $|\psi_z^{(0)}\rangle$ and $|\psi_y^{(0)}\rangle$. The larger matrix elements of $H_{\text{HF}\perp}$ involve states separated by the large zero-field splitting (2870 MHz) and thus introduce negligible mixing $\sim A_\perp/D$, but mixing due to the smaller matrix elements cannot be neglected, since the states are separated only by the nuclear quadrupolar frequency (~5 MHz). In particular, the mixing of states separated by the quadrupolar frequency is $\sim \gamma_e B_0 \cdot A_\perp/\left(D \cdot P\right)$, while the mixing of states separated by the zero-field splitting is smaller by a factor of ~40. Note that the significant mixing of states is a second-order effect that depends on two nonsecular terms in the Hamiltonian.

A resonant MW field induces allowed transitions (marked as solid arrows in Fig. 3) but also causes additional forbidden transitions (marked as dashed arrows in Fig. 3), due to this second-order effect. The spin coherences that develop as a result of the forbidden transitions involve linear combinations of energy eigenstates that have the nuclear spin in different eigenstates of the quadrupolar Hamiltonian, and the spin-echo envelope is modulated at the quadrupolar frequency because of the evolution of these linear combinations. The coherences are not fully refocused by the MW π pulse unless the period of free evolution that precedes the π pulse allows for an integral number of oscillations at the quadrupolar frequency.

## IV. RESULTS AND DISCUSSIONS

For quantitative analysis, numerical simulations were carried out using the density matrix formalism. The spin Hamiltonian shown in Eq. (1) was first numerically diagonalized. Simulations were performed in an interaction frame where the large energy differences were removed from the diagonalized Hamiltonian. The time-dependent Hamiltonian $\gamma_e B_1 \cdot S_x \mathrm{Cos}(\omega t)$, which represents a resonant MW field with frequency $\omega$ and amplitude $B_1$ directed along the $x$-axis, was first represented in the energy eigenbasis and then transformed into this interaction frame for simulation of the pulses.

The population of the electron spins after optical polarization was assumed to be in state $|m_s = 0\rangle$. After the MW $\pi/2$ pulse, spin coherences freely evolve under the interaction-frame

Hamiltonian during $t = 2\tau$, with a $\pi$ pulse applied at $t = \tau$. At $t = 2\tau$, another $\pi/2$ pulse is applied to convert coherences to populations, and the ESEEM signal for each $\tau$ is obtained from the resulting population in state $|m_s = 0\rangle$.

For numerical simulations, we used parameters corresponding to the experimental conditions: $E/2\pi = 2.75$ MHz, $B_0 = 75$ G, $\gamma_e B_1/2\pi = 5.00$ MHz. Coherence decay due to the various spin-spin relaxation processes was phenomenologically included in the simulation. As shown in Fig. 1, the simulations showed excellent agreement with experimental results. When the magnetic field was applied along the $x$-axis, perpendicular to the NV axis, observable modulation appeared in the ESEEM signal. No modulation was observed when the magnetic field was applied along the $z$ axis, parallel to the NV quantization axis. The simulated modulation depth (~4%) of the ESEEM signal also showed excellent agreement with experimental data. When a CPMG pulse sequence with 10 $\pi$ pulses was employed in the experiments, this modulation was amplified, as shown in the inset of Fig. 1.

The frequency spectra shown in Fig. 4 were obtained by taking the Fourier transform of the experimental and simulation data after exponential decay was subtracted from the data. (The result of subtracting the exponential decay is shown in the inset of Fig. 4.) The experimental spectrum acquired with the Hahn spin-echo sequence showed excellent quantitative agreement with the simulated spectrum at ~5 MHz. Simulation and experimental data also showed a spectral peak at ~10 MHz, consistent with the fact that ESEEM gives envelope modulation at frequencies $\omega_0$, $\omega_+$, $\omega_0 - \omega_+$, and $\omega_0 + \omega_+$, where $\omega_0$ and $\omega_+$ stand for the transition frequencies of the nuclear spin associated with the electron spin states $|\psi_z\rangle$ and $|\psi_y\rangle$, respectively.[26]

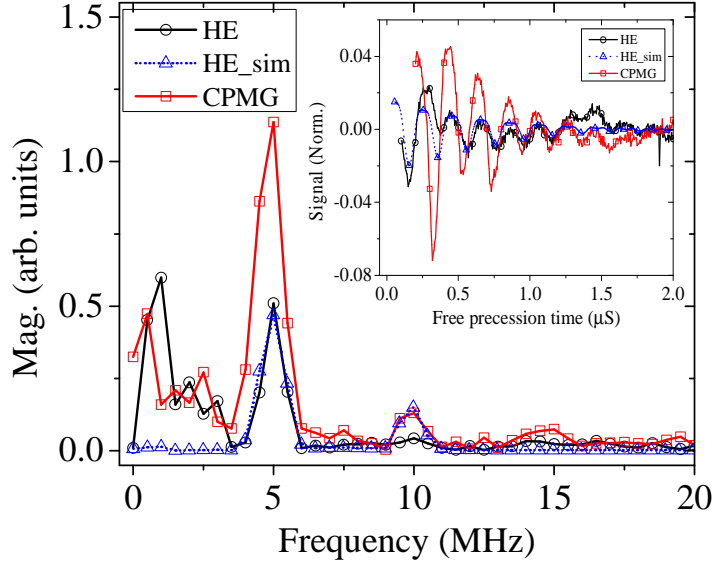

**FIG. 4.** (Color online) Fourier-transform spectra of the ESEEM signal of the NV center in an external magnetic field of 75 G applied perpendicular to the NV axis. Solid lines with circles or squares represent experimental data obtained using the Hahn echo (HE) pulse sequence or the CPMG pulse sequence, respectively. The dotted line with triangles represents a simulation that used the Hahn echo pulse sequence. The inset shows the ESEEM signal after the exponential decay was subtracted.

It was confirmed in the simulation that $H_{HF\perp}$ is responsible for the electron-nuclear mixing, since the modulation disappeared when $A_\perp$ was set to zero. It was also confirmed that the frequencies of modulation changed linearly with the quadrupole coupling frequency. Under our experimental conditions, the modulation frequencies of the spin-echo envelope are primarily determined by the quadrupolar interaction of the $^{14}$N nuclear spin. The energy eigenstates responsible for the modulation can be approximated as product states in which the nuclear spin is in an eigenstate of the quadrupolar Hamiltonian $H_Q$, and $H_Q$ makes the dominant contribution to the energy differences that determine the modulation frequencies; in particular, the contribution form the hyperfine interaction and the Zeeman interaction is less than 1 % of the contribution of the quadrupolar interaction.

Other frequency components in the experimental data could originate from the hyperfine interactions with neighboring $^{13}$C nuclei in the diamond lattice. We note that the modulation frequencies due to the hyperfine interaction with $^{13}$C nuclei could be greatly modified from the secular term $A_{zz}$ of the hyperfine interaction, since the net contribution of the hyperfine interaction to transition frequencies is a function of magnetic field strength and orientation when a static magnetic field is misaligned from the NV quantization axis. This has been well studied in the literature and is not a focus of our discussion.[26,37,38]

In our method, the frequency shift of the ESEEM signal due to the Zeeman Hamiltonian is less than ~1 %, so the measurement of the quadrupole interaction does not require a highly accurate measurement of the applied field. The accuracy of our measurement method is currently limited by the NV spin coherence time, since the oscillation due to the quadrupole interaction is indirectly measured by observation of the electron spin coherence. Therefore with a high-purity diamond sample that has a lower concentration of NV centers and $^{14}$N impurities, the measurement accuracy can be improved.

Finally, we also note that the spin coherence time was enhanced by a factor of ~2, when the magnetic field was oriented perpendicular to the NV quantization axis, as shown in Fig. 1. This was attributed to the fact that the effective noise amplitude can be renormalized by the combined effect of the transverse magnetic field and the zero-field splitting parameters, as discussed in our recent paper.[42]

## V. CONCLUSION

We reported optical detection of the $^{14}$N nuclear quadrupole coupling in the NV center by employing ESEEM techniques in the presence of a transverse magnetic field. Numerical simulations and a theoretical model showed excellent agreement with experimental results. By applying a small magnetic field of 75 G perpendicular to the NV quantization axis, we modified the symmetry of the spin system. The $^{14}$N nuclear quadrupole coupling, which is normally undetectable in spin-echo experiments with the NV center, modulated the spin-echo envelope as a result of forbidden transitions associated with second-order mixing involving two nonsecular terms in the Hamiltonian. Although other techniques such as optically detected Raman heterodyne ENDOR and optically detected Raman heterodyne NMR have been employed to detect the quadrupolar interaction, our technique is experimentally much simpler. In principle, this technique can be used to map out the hyperfine tensor and/or the nuclear quadrupole coupling tensor of a spin system with a similar energy structure.[24]

This work was supported by the Director, Office of Science Office of Basic Energy Sciences, Materials Sciences and Engineering Division, of the US Department of Energy under Contract No. DE-AC02-05CH11231. M. B. acknowledges salary support from the National Science Foundation under award number CHE-0957655. R. L. acknowledges salary support from Hong Kong RGC-NSFC Project N_CUHK403/11.

*Current address: Halliburton Energy Services, Inc., 3000 N Sam Houston Pkwy. E, Houston, TX 77032, USA.

†Current address: William R. Wiley Environmental Molecular Sciences Laboratory, Pacific Northwest National Laboratory, Richland, Washington 99352, USA

‡vikbajaj@gmail.com

# REFERENCES

[1] J. Harrison, M.. Sellars, and N.. Manson, J. Lumin. **107**, 245 (2004).

[2] G. Balasubramanian, P. Neumann, D. Twitchen, M. Markham, R. Kolesov, N. Mizuochi, J. Isoya, J. Achard, J. Beck, J. Tissler, V. Jacques, P.R. Hemmer, F. Jelezko, and J. Wrachtrup, Nat. Mater. **8**, 383 (2009).

[3] G. Balasubramanian, I.Y. Chan, R. Kolesov, M. Al-Hmoud, J. Tisler, C. Shin, C. Kim, A. Wojcik, P.R. Hemmer, A. Krueger, T. Hanke, A. Leitenstorfer, R. Bratschitsch, F. Jelezko, and J. Wrachtrup, Nature **455**, 648 (2008).

[4] J.M. Taylor, P. Cappellaro, L. Childress, L. Jiang, D. Budker, P.R. Hemmer, A. Yacoby, R. Walsworth, and M.D. Lukin, Nat. Phys. **4**, 810 (2008).

[5] C. Shin, C. Kim, R. Kolesov, G. Balasubramanian, F. Jelezko, J. Wrachtrup, and P.R. Hemmer, J. Lumin. **130**, 1635 (2010).

[6] N. Zhao, J. Honert, B. Schmid, M. Klas, J. Isoya, M. Markham, D. Twitchen, F. Jelezko, R.-B. Liu, H. Fedder, and J. Wrachtrup, Nat. Nanotechnol. (2012).

[7] S. Kolkowitz, Q.P. Unterreithmeier, S.D. Bennett, and M.D. Lukin, Phys. Rev. Lett. **109**, 137601 (2012).

[8] G. de Lange, D. Ristè, V.V. Dobrovitski, and R. Hanson, Phys. Rev. Lett. **106**, 080802 (2011).

[9] H.J. Mamin, M. Kim, M.H. Sherwood, C.T. Rettner, K. Ohno, D.D. Awschalom, and D. Rugar, Science **339**, 557 (2013).

[10] T. Staudacher, F. Shi, S. Pezzagna, J. Meijer, J. Du, C.A. Meriles, F. Reinhard, and J. Wrachtrup, Science **339**, 561 (2013).

[11] J.A.S. Smith, J. Chem. Educ. **48**, 39 (1971).

[12] C.P. Slichter, *Principles of Magnetic Resonance (Springer Series in Solid-State Sciences)* (Springer, 1996).

[13] E. Balchin, D.J. Malcolme-Lawes, I.J.F. Poplett, M.D. Rowe, J.A.S. Smith, G.E.S. Pearce, and S.A.C. Wren, Anal. Chem. **77**, 3925 (2005).

[14] V.S. Grechishkin and N.Y. Sinyavskii, Phys.-Uspekhi **40**, 393 (1997).

[15] J.A.S. Smith, J. Chem. Educ. **48**, A243 (1971).

[16] E.G. Brame, Anal. Chem. **39**, 918 (1967).

[17] H.D. Schultz and C. Karr, Anal. Chem. **41**, 661 (1969).

[18] J. Schmidt and J.H. Van der Waals, Chem. Phys. Lett. **3**, 546 (1969).

[19] K.P. Dinse and C.J. Winscom, J. Chem. Phys. **68**, 1337 (1978).

[20] Y. k. Lee, Concepts Magn. Reson. **14**, 155 (2002).

[21] M.D. Hürlimann, C.H. Pennington, N.Q. Fan, J. Clarke, A. Pines, and E.L. Hahn, Phys. Rev. Lett. **69**, 684 (1992).

[22] S.-K. Lee, K.L. Sauer, S.J. Seltzer, O. Alem, and M.V. Romalis, Appl. Phys. Lett. **89**, 214106 (2006).

[23] L.G. Rowan, E.L. Hahn, and W.B. Mims, Phys. Rev. **137**, A61 (1965).

[24] D.J. Singel, W.A.J.A. van der Poel, J. Schmidt, J.H. van der Waals, and R. de Beer, J. Chem. Phys. **81**, 5453 (1984).

[25] M.K. Bowman and J.R. Norris, J. Chem. Phys. **77**, 731 (1982).

[26] V. Weis, K. Möbius, and T. Prisner, J. Magn. Reson. **131**, 17 (1998).

[27] N.B. Manson, X.-F. He, and P.T.H. Fisk, Opt. Lett. **15**, 1094 (1990).

[28] N.B. Manson, X.-F. He, and P.T.H. Fisk, J. Lumin. **53**, 49 (1992).

[29] E.L. Hahn, Phys. Rev. **80**, 580 (1950).

[30] H.Y. Carr and E.M. Purcell, Phys. Rev. **94**, 630 (1954).

[31] S. Meiboom and D. Gill, Rev. Sci. Instrum. **29**, 688 (1958).

[32] C.S. Shin, C.E. Avalos, M.C. Butler, D.R. Trease, S.J. Seltzer, J. Peter Mustonen, D.J. Kennedy, V.M. Acosta, D. Budker, A. Pines, and V.S. Bajaj, J. Appl. Phys. **112**, 124519 (2012).

[33] J.H.N. Loubser and J.A. van Wyk, Rep. Prog. Phys. **41**, 1201 (1978).

[34] A. Gali, M. Fyta, and E. Kaxiras, Phys. Rev. B **77**, 155206 (2008).

[35] N.B. Manson, J.P. Harrison, and M.J. Sellars, Phys. Rev. B **74**, 104303 (2006).

[36] E. Van Oort and M. Glasbeek, Chem. Phys. **143**, 131 (1990).

[37] L. Childress, M.V.G. Dutt, J.M. Taylor, A.S. Zibrov, F. Jelezko, J. Wrachtrup, P.R. Hemmer, and M.D. Lukin, Science **314**, 281 (2006).

[38] N. Zhao, J.-L. Hu, S.-W. Ho, J.T.K. Wan, and R.B. Liu, Nat. Nanotechnol. **6**, 242 (2011).

[39] P.L. Stanwix, L.M. Pham, J.R. Maze, D. Le Sage, T.K. Yeung, P. Cappellaro, P.R. Hemmer, A. Yacoby, M.D. Lukin, and R.L. Walsworth, Phys. Rev. B **82**, 201201 (2010).

[40] X.-F. He, N.B. Manson, and P.T.H. Fisk, Phys. Rev. B **47**, 8816 (1993).

[41] R. Hanson, V.V. Dobrovitski, A.E. Feiguin, O. Gywat, and D.D. Awschalom, Science **320**, 352 (2008).

[42] C.S. Shin, C.E. Avalos, M.C. Butler, H.-J. Wang, S.J. Seltzer, R.-B. Liu, A. Pines, and V.S. Bajaj, Phys. Rev. B **88**, 161412 (2013).